\documentclass[12pt,preprint]{aastex}

\slugcomment{}

\shorttitle{\nh3 Observations of the Infrared Dark Cloud
G28.34+0.06} \shortauthors{Wang et al.}

\newcommand{\skipthis}[1]{}
\def\nh3{$\rm{NH_3}$}
\def\NH3{$\rm{NH_3}$}
\def\msolar{M$_\odot$}
\def\msun{M$_\odot$}

\def\lsun{L$_\odot$}
\def\kms-1{km~s$^{-1}$}
\def\h2{$\rm{H_2}$}

\def\cm-3{$\rm{cm^{-3}}$}

\begin{document}


\title{\nh3 Observations of the Infrared Dark Cloud G28.34+0.06}


%\author{Yang Wang$^{1,2}$, Qizhou Zhang$^{2}$, Jill M. Rathborne$^{3}$, James Jackson$^{3}$, Yuefang Wu$^{1}$,}

%\author{\\}
%\affil{}

\author{Y. Wang$^{1,2}$, Q. Zhang$^{2}$, T. Pillai$^{2,3}$, F. Wyrowski$^3$, Y. Wu$^{1}$,}

\altaffiltext{1}{Astronomy Department and CAS-PKU Joint Beijing
Astrophysics Center, Peking University, Beijing
   100871, P.R.China}

\altaffiltext{2}{Harvard-Smithsonian Center for Astrophysics,
Cambridge, MA 02138, U.S.A.}

\altaffiltext{3}{Max-Planck-Institut f{\"u}r Radioastronomie, Auf
dem H{\"u}gel 69, D-53121 Bonn, Germany}

\begin{abstract}

We present observations of the \nh3 (J,K) = (1,1) and (2,2)
inversion transitions toward the infrared dark cloud G28.34+0.06,
using the Very Large Array. Strong \nh3 emission is found to
coincide well with the infrared absorption feature in this cloud.
The northern region of G28.34+0.06 is dominated by a compact clump
(P2) with a high rotation temperature (29 K), large line width
(4.3 km s$^{-1}$), and is associated with strong water maser (240
Jy) and a 24 $\mu$m point source with far IR luminosity of $10^3$
\lsun. We infer that P2 has embedded massive protostars although
it lies in the 8 $\mu$m absorption region. The southern region has
filamentary structures. The rotation temperature in the southern
region decreases with the increase of the integrated \nh3
intensity, which indicates an absence of strong internal heating
in these clumps. In addition, the compact core P1 in the south has
small line width (1.2 km s$^{-1}$) surrounded by extended
emission with larger line width (1.8 km s$^{-1}$), which suggests
a dissipation of turbulence in the dense part of the cloud. Thus,
we suggest that P1 is at a much earlier evolutionary stage than
P2, possibly at a stage that begins to form a cluster with massive stars.

\end{abstract}


\section{Introduction}

Stars form from the collapse of dense molecular gas and dust.
While low-mass stars tend to form in isolation and small groups,
high-mass stars ( M $>$ 8 \msun) are born almost exclusively in
clusters (Lada \& Lada 2003). The process of low-mass star
formation in the isolated environment is relatively well studied
(Shu, Adams \& Lizano 1987). However, the processes through which
gas and dust condense to assemble massive stars and clusters are
poorly understood. Klessen (2004) and Mac Low (2004) proposed
that, in large scales, the supersonic turbulence provides the
support to the cloud, while, in small scales, density enhancements
cause gravitational collapse. And Myers (1998) and 
Goodman et al. (1998) also proposed that,
after the turbulence dissipation, gas condensation will loss
internal support against the gravity to form stars. Such scenario
requires observational tests in massive regions where clouds are
very likely to form massive stars and clusters. Studies
of massive infrared dark clouds (IRDCs) can provide such opportunities.

Infrared dark clouds are the dark patches in the sky revealed in
the infrared wavelengths against the bright galactic background.
Surveys along the galactic plane using the Infrared Space
Observatory (ISO) and the Midcourse Space Experiment (MSX)
identified thousands of extinction features at 8 $\mu$m (Egan et
al. 1998; Hennebelle et al. 2001; Simon et al. 2006). Their
spatial coincidence with emission from molecular lines and dust
indicates that these dark clouds consist of dense molecular gas
and dust of $n_{H_2} \sim 10^5$ cm$^{-3}$ (Carey et al. 1998, 2000).
Temperatures in these clouds are low enough (10--20 K) that they
do not radiate significantly even at the mid or far infrared
wavelength. The large amount of dense gas and dust in these clouds
make them the natural birth place for clusters (Carey et al. 1998,
2000; Johnstone et al. 2003; Rathborne et al. 2005, 2006, 2007;
Teyssier, Hennebelle \& P{\'e}rault 2002; Redman et al. 2003;
Pillai et al. 2006a,b; Sridharan et al. 2005; Beuther et al.
2005).

However, most of the previous studies focus on IRDCs that already
have strong star formation activities (e.g. Rathborne et al.
2005), and thus, are at a more evolved evolutionary stage. In
addition, due to the large distances to these objects,
observations with single dish telescopes do not have sufficient
spatial resolution to reveal the substructures in massive clumps
and variations of physical parameters. Using the Very Large Array,
we imaged a giant infrared dark cloud, G28.34+0.06 at high angular
resolution. The object of interest is a good target to search for
molecular clumps in an early evolutionary stage. At a distance of
$\sim$4.8 kiloparsecs (kpc), it contains about 10$^3$ M$_\odot$ of
dense gas in the infrared absorption region (Carey et al. 2000;
Rathborne et al. 2006). The \nh3 observations with the Effelsberg
100m telescope show that, on average, its line width is smaller
than those obtained in high mass protostellar objects and
Ultracompact HII (UCHII) regions, and the gas temperature is lower
than 20 K (Pillai et al. 2006b). In our VLA observations, the
temperature structure and line width variations indicate that the
northern region of G28.34 harbors massive protostellar objects,
while, in the southern region, the clumps have characteristics of
the earliest phase of star formation.

\section{Observations}

We observed the \NH3 (J, K) = (1,1) line at 23.694 GHz and the
(2,2) line at 23.723 GHz with the VLA of the NRAO\footnote{The
National Radio Astronomy Observatory is operated by Associated
Universities, Inc., under cooperative agreement with the National
Science Foundation.} on 2004 June 15 in its D configuration. Seven
pointings were observed to cover the extended emission; each
pointing has an on-source integration time of 12 minutes and covers a
FWHM primary beam of 2$'$. We used the 4IF mode that splits the
256-channel correlator into four sections to allow simultaneous
observations of the \NH3 (1,1) and (2,2) lines with two circular
polarizations for each line. The channel separation used was 48.8
KHz ($\sim$0.6 km s$^{-1}$ at the line frequency). The time
variation of the gains was calibrated by quasar, 1851+005,
observed at a cycle of $\sim$20 mins. The absolute flux density is
established by bootstrapping to 3C286. The bandpass is calibrated
via observations of 3C84.

The visibility data sets were calibrated and imaged using the AIPS
software package of the NRAO. The average rms of the final images
is $\sim$ 3 mJy/beam per 48.8 KHz wide channel. The synthesis beam
is about 3$''$$\times$5$''$ when using the natural weighting of
the visibilities. In order to recover extended structures missing
in the interferometer data, we combined the visibility data from
the VLA with the Effelsberg 100m single dish data from Pillai et
al. (2006b) for both the \NH3 (1,1) and (2,2) lines following the
Miriad procedure outlined in Vogel et al. (1984).

\section{Results and Discussions}

\subsection{General Properties}

Fig. 1 presents the integrated intensity of the \NH3 (1,1) line
from the combined data set overlaid on the 8 $\mu$m image obtained
from the Galactic Legacy Infrared Mid-Plane Survey Extraordinaire
(GLIMPSE; PI: Churchwell) using the Infrared Array Camera (IRAC).
The dense gas traced by \nh3 follows the 8 $\mu$m absorption
feature very well. Several compact clumps are revealed in the
filamentary structure. These clumps also coincide with the 1.2 mm
dust continuum emission obtained from the IRAM 30m telescope
(Rathborne et al. 2006).

In the region north of $\delta$$_{(2000)}$ $<$ -04:00:45
(hereafter the Northern Region), both the \nh3 and 1.2 mm
emission are centrally peaked with a dust mass of $\geq$600
\msun\ (Carey et al. 2000). There exists an HII region associated
with IRAS 18402-0403, but has little \nh3 or 1.2 mm emission. The
24 $\mu$m image from the 24 and 70 Micron Survey of the Inner
Galactic Disk with MIPS (MIPSGAL; PI: Carey) reveals five embedded
point sources. IRS 3 coincides with IRAS 18402-0403, and has a
luminosity of 10$^3$ \lsun (Carey et al. 2000). IRS 2 is detected
toward the 1.2mm continuum peak P2, and is associated with the
\nh3 peak and a strong H$_2$O maser (240 Jy). Its luminosity of
10$^3$ \lsun\ based on the Spitzer data from 3.6 to 70 $\mu$m and
1.2mm flux, and a lack of cm emission suggest a
high-mass protostellar object. The remaining three 24 $\mu$m
sources have luminosities of $10^2$ \lsun.

The region south of $\delta$$_{(2000)}$ $<$ -04:00:45 (hereafter
the Southern Region) has no detectable emission at 1.3cm at an rms
of 0.7 mJy, and is associated with three H$_2$O masers (Wang et
al. 2006). Both the \nh3 and 1.2mm emission have a flatter
spatial distribution, with a total mass over 1500 \msun\
(Rathborne et al. 2006). Four 24 $\mu$m sources are detected from
the MIPSGAL survey with luminosities  $ < 10^2$ \lsun. The
strongest source, IRS 1 ($\alpha$$_{(2000)}$=18:42:50.82
$\delta$$_{(2000)}$= -04:03:09) with a luminosity of $10^2$ \lsun,
is associated with an \nh3 peak toward P1
($\alpha$$_{(2000)}$=18:42:50.27 $\delta$$_{(2000)}$= -04:03:20),
and an H$_2$O maser emission ($\alpha$$_{(2000)}$=18:42:50.80
$\delta$$_{(2000)}$= -04:03:11.0). The other two H$_2$O masers in
the Southern Region has no 24 $\mu$m sources within a projected
distance of 30$''$.

{\bf It is difficult to predict whether the Southern Region will 
eventually form massive stars similar to the Northern Region.}
The Southern Region contains more dense molecular gas than the 
Northern Region. Assuming a similar star forming efficiency, one
would expect a more massive cluster to emerge in the Southern
Region. {\bf However, since the south is spatially more enxtended, it may form
less massive stars than the Northern Region. On the other hand, if the cluster
follows a normal Salpeter IMF, one would expect more massive stars in
the Southern Region.} 
Thus, a lack of luminous infrared sources in the south
indicates that it is at an earlier evolutionary stage than
the Northern Region. Furthermore, the radio data in conjunction
with the infrared data appear to suggest an evolutionary sequence
of star formation, starting with an evolved stage in
IRAS 18402-0403/IRS3; to an embedded phase in IRS2 which has a
luminosity similar to high-mass protostellar objects; to less
luminous IRS1 which has strong mm continuum, and H$_2$O maser emission;
and finally to \nh3 clumps which only has strong mm
continuum emission, but no 24 $\mu$m emission peak nor H$_2$O
maser emission.

The \nh3 data reveal physical properties that further confirm
the difference in the Northern and Southern Region. In the
Northern Region, there is a dominant compact clump P2 associated
with IRS 2 ($\alpha$$_{(2000)}$=18:42:51.92 $\delta$$_{(2000)}$=
-03:59:54), which has a peak rotation temperature close to 30 K
and a FWHM line width larger than 3 km s$^{-1}$, and is surrounded by an
extended envelope. By averaging rotation temperatures in the
Northern Region of the same \nh3 (1,1) intensities, we derived a
relation between the two quantities.
Fig. 2 presents the average \nh3 temperature within an
intensity bin versus the mean value of the bin. As shown in the
black line in Fig. 2, the rotation temperature in the Northern
Region increases with the increase of the integrated intensity,
which indicates strong internal heating in clump P2 of the
Northern Region. Therefore, although clump P2 is embedded
in the 8 $\mu$m dark region, the high temperature, large line
width, strong water maser, and 24 $\mu$m point source indicate that
this compact clump is forming massive protostellar object(s),
similar to the compact cores found in other IRDCs (Pillai et al.
2006a; Rathborne et al. 2005, 2007; Beuther et al. 2005).

On the other hand, the \nh3 emission from the Southern Region
appears to have different characteristics as compared to that in
the Northern Region. The red line in Fig. 2 shows that the
rotation temperature in the Southern Region decreases with the
increase of the integrated \nh3 intensity. On average, areas with
stronger \nh3 emission have lower temperatures, which indicates absence
of strong internal heating in the southern clumps, unlike P2 in the
north. Toward the most compact clump P1/IRS1 in the Southern
Region, the \nh3 gas has a lower rotation temperature ($\sim$16 K)
and narrower line width ($\sim$1.8 km s$^{-1}$) than that in P2.
We also note that the temperature decrease appears to flatten
toward the highest intensity in the red line in Fig. 2. This could
be due to the limited heating from star formation activities in
the area.

We also compared the general properties of G28 Southern Region
with the local, intermediate mass star formation region OMC-north
(Bally et al. 1987; Hillenbrand 1997; Castets \& Langer 1995;
Johnstone \& Bally 1999). The linear resolution of 
our interferometer and single dish
combined \nh3 image ($\sim$4$''$ resolution) of G28.34+0.06
(distance of $\sim$4.8 kpc) is comparable with the Effelsberg
100m single dish \nh3 image (40$''$ resolution) of the OMC-north
(distance of $\sim$0.48 kpc) (Cesaroni et al. 1994).
We found that, on average, G28.34 Southern
Region have lower temperatures (16 K vs. 20 K), larger line width
(1.6 vs 0.8 km s$^{-1}$) and higher column density (2.3 vs.
0.78$\times$10$^{22}$ cm$^{-2}$) assuming the same \nh3 abundance
(0.3 $\times$ 10$^{-7}$) in both regions. The mass within the
dense filament estimated from dust emission is $\sim$1500 \msolar\
in G28.34-south region (Rathborne et al. 2006), while the dense
mass in OMC-north is $\sim$350 \msolar\ (Lis et al. 1998). On the
other hand, G28.34-south region extends 4$'$ ($\sim$6 pc) at the
distance of $\sim$4.8 kpc, while OMC-north extends 20$'$($\sim$3
pc) at the distance of $\sim$0.48 kpc. All these parameters
indicate that G28.34-south has colder and more turbulent dense gas
packed in a smaller region than OMC-north.

\subsection{Line Width Variation of \nh3 Emission toward Peaks P1 and P2}

We examined the spatial variation of \nh3 (1,1) line width from
the VLA data as well as that combined with the Effelsberg
100m single dish data. The line width variations after the
data combination provide kinematic information on the scale of
$\geq$ 0.1 pc (4$''$ in 4.8 kpc). The line width of P1 and P2
before and after the data combination are shown in Table 1. It
appears that the line width in P1 is broader after adding the
single dish data, while, in P2, we find an opposite effect. Since
the missing flux in an interferometer corresponds to the extended
spatial structures, the variation in the line width represents a
change in the level of turbulence in their envelopes/cores along
the line of sight. Normally, in a typical active massive star
forming core, such as P2, the missing fluxes in an interferometer
are from extended and quiescent envelopes. Dense and compact
emission surrounding high mass protostars or UCHII regions has
broader line width due to motions such as infall, outflow and
rotation. The missing extended structures are normally weaker and
have narrower line width. However, toward the G28.34 Southern
Region, we observed the opposite effect: the extended emission
shows systematically larger line widths than those from the
compact core component revealed by interferometer.

This effect is also shown in line width distribution in Fig. 3.
The line width distribution appears to behave differently in the
clumps from the Northern and Southern region after data
combination. As shown in Fig. 3, for the cold clump P1 in the
Southern Region, the mean line width becomes broader after
recovering the missing flux, while for the warm clump P2, the mean
line width is similar to the one from the interferometer data
alone. Furthermore, the line widths of the entire Southern Region
follow a distribution similar to P1. We find that the effect of
relative motions among clumps can be ruled out. We tested this
scenario by convolving the combined interferometer data to the
resolution of the single dish telescope (40$''$), and
comparing with the line width of the combined data at $4''$
resolution. The mean line widths from these two data sets are
similar.

The variations of \nh3 line width in the G28 Southern Region
suggest a dissipation of turbulence, possibly at the onset of
star/cluster formation. The dissipation of turbulence could
initiate gravitational collapse as suggested by several
observations (e.g. Ward-Thompson, D., 2002). Taking advantage of
the high resolution of interferometers, we obtained the line width
variation of P1 and P2 along the line of sight on scales of
$\geq$0.1 pc.  Using the \nh3 data, we calculated the molecular
mass and virial mass (Estalella et al. 1993) of P1 and P2 (Table
1). In P1, the material seen from both interferometer alone and
the combined data sets is gravitationally bound. And the dense
component is more likely to collapse (M = 16 M$_\odot$ vs. M$_{vir}$
= 6 M$_\odot$) than the component including all the spatial
structures (M = 27 M$_\odot$ vs. M$_{vir}$ = 14 M$_\odot$). In
contrast, due to strong star formation activities, the dense
component in P2 has a larger virial mass (75 M$_\odot$) than the
molecular mass (30 M$_\odot$), while the whole clump is
gravitationally bound and may undergo collapse (M = 75 M$_\odot$
vs. M$_{vir}$ = 44 M$_\odot$). We suggest that, in the cold
massive clump P1, and the rest of the G28.34-south region,
dense gas undergoes fragmentation and condensation. Considering
the low rotation temperature ($\sim$16 K), relatively narrow line
width ($\sim$1.8 km s$^{-1}$), P1 appears to be in a very early
evolutionary stage during which turbulence dissipates to facilitate
gravitational collapse to form stars. At this stage of
cluster/massive star formation, the most massive members in the
cluster may be still at the intermediate mass stage, and will likely
continue to accrete to become more massive stars, given the amount
of dense gas available in the region.

\begin{deluxetable}{cccccccccc}
\tabletypesize{\scriptsize}
%\rotate

\tablecaption{Physical Parameters of P1 and P2 over the Scale of
0.1 pc}

\tablewidth{0pt}

\tablehead{

\colhead{Source} & \colhead{R.A. (2000)} & \colhead{DEC. (2000)}&
\colhead{$\Delta\upsilon$$^a$} & \colhead{N$_{col}$$^a$} &
\colhead{Mass$^a$}
&\colhead{Virial Mass$^a$}&\colhead{\nh3 Abundance$^b$} \\

\colhead{Name} & \colhead{} & \colhead{}  &  \colhead{(km
s$^{-1}$)} & \colhead{(10$^{16}$cm$^{-2}$)} &
\colhead{(M$_\odot$)} & \colhead{(M$_\odot$)}&
\colhead{(10$^{-7}$)}

}

\startdata

P1$^c$  & 18 42 50.27  & -04 03 20  &  $1.2/1.8$ & $0.59/1.0$  & 16/27 & 6/14 & 0.6 \\
P2$^c$  & 18 42 52.07  & -03 59 54  &  $4.3/3.3$ & $0.48/1.2$  & 30/75 & 75/44 & 0.3 \\

\enddata

\tablecomments{Units of right ascension are hours, minutes, and
seconds, and units of declination are degrees, arcminutes, and
arcseconds.}

\tablenotetext{a}{The first number is obtained from the VLA data
only; The second number is obtained from the combined data.}

\tablenotetext{b}{The \nh3 abundance is calculated by comparing
the column density of dust (Rathborne et al. 2006) and \nh3
emission. We convolved the \nh3 data to $11''$, the the resolution
of the dust emission, and assumed that the \nh3 abundance is the
same in ($11'' \times 11''$).}

\tablenotetext{c}{The coordinates of P1 and P2 refer to the
integrated intensity peaks of combined \nh3 (1,1) data.}

\end{deluxetable}

\acknowledgments

We thank referee D. Johnstone for his constructive comments.
This research is supported by the Grant 10128306 and 10733030 of
NSFC and G1999075405 of NKBRSF. This work is based in part on
observations made with the {\it Spitzer Space Telescope}, which is
operated by the Jet Propulsion Laboratory, California Institute of
Technology. We would like to thank the GLIMPSE team (PI: E.
Churchwell) and  MIPSGAL team (PI: S. Carey) for making the IRAC
and MIPS images available to the community. The authors also
gratefully acknowledge Mr. Keping Qiu's help in reducing the MIPS
data.

%\section{Captions of figures}

\newpage

\begin{figure}[h]
%\begin{center}

\includegraphics{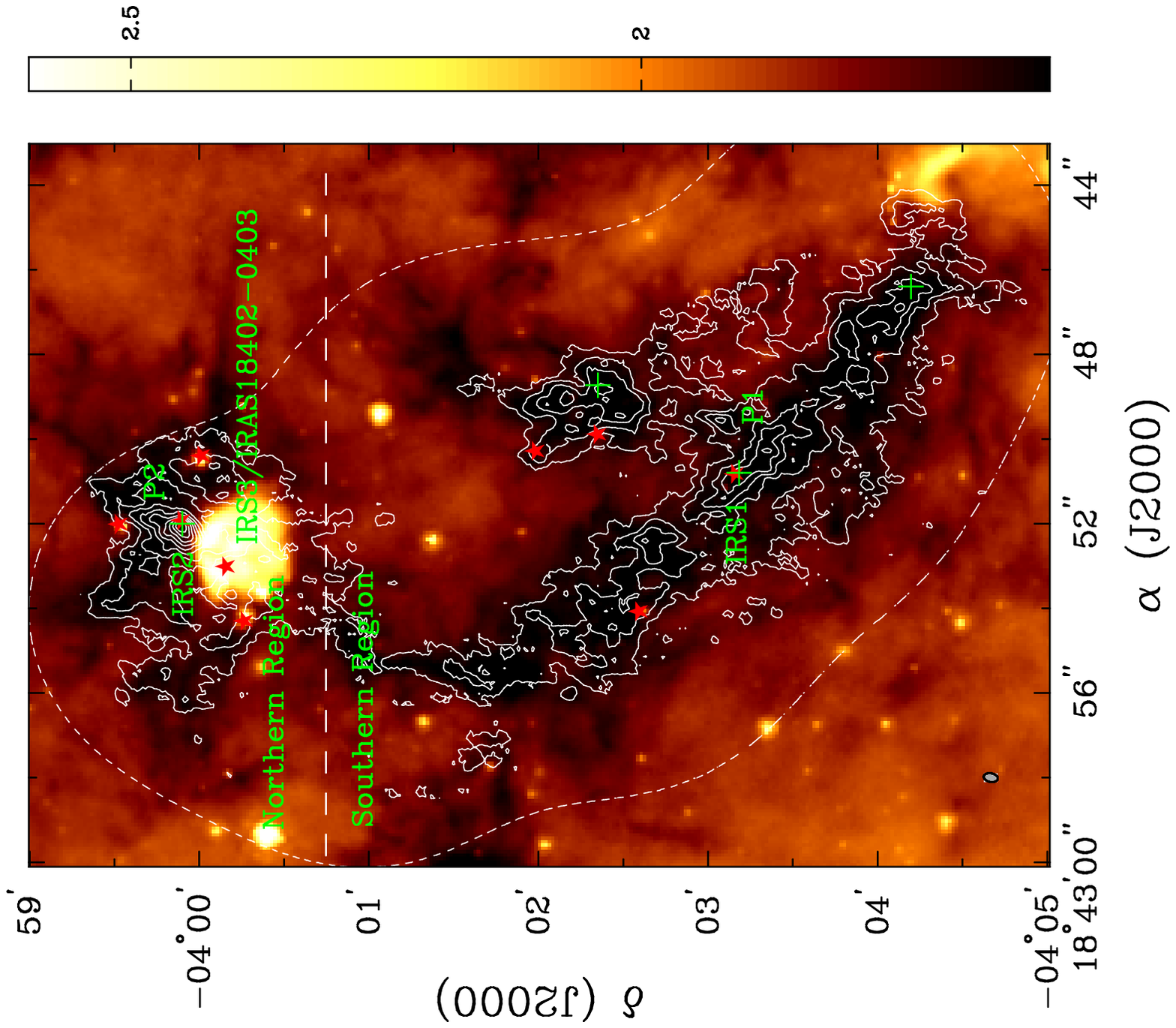}

\caption{ The integrated intensity of the combined \nh3 (1,1)
emission in white solid contours overlaid on the Spitzer 8 $\mu$m
image in logarithmic color scales. The \nh3 image is contoured at
10\% of the peak (1 Jy beam$^{-1}$ $\times$ km s$^{-1}$). The star
symbols mark the 24 $\mu$m emission peaks observed with
MIPS/\emph{Spitzer}. The cross symbols mark water maser
emission detected with the VLA. The thin dashed line indicates the
50\% of the sensitivity level of the 7 pointing mosaic in \nh3.
The \nh3 data have a resolution of $5'' \times 3''$ , shown as the
shaded ellipse at the lower-left corner of the panel. }

\end{figure}

\newpage

%\vskip 3.5in

\begin{figure}[h]
%\begin{center}

\includegraphics{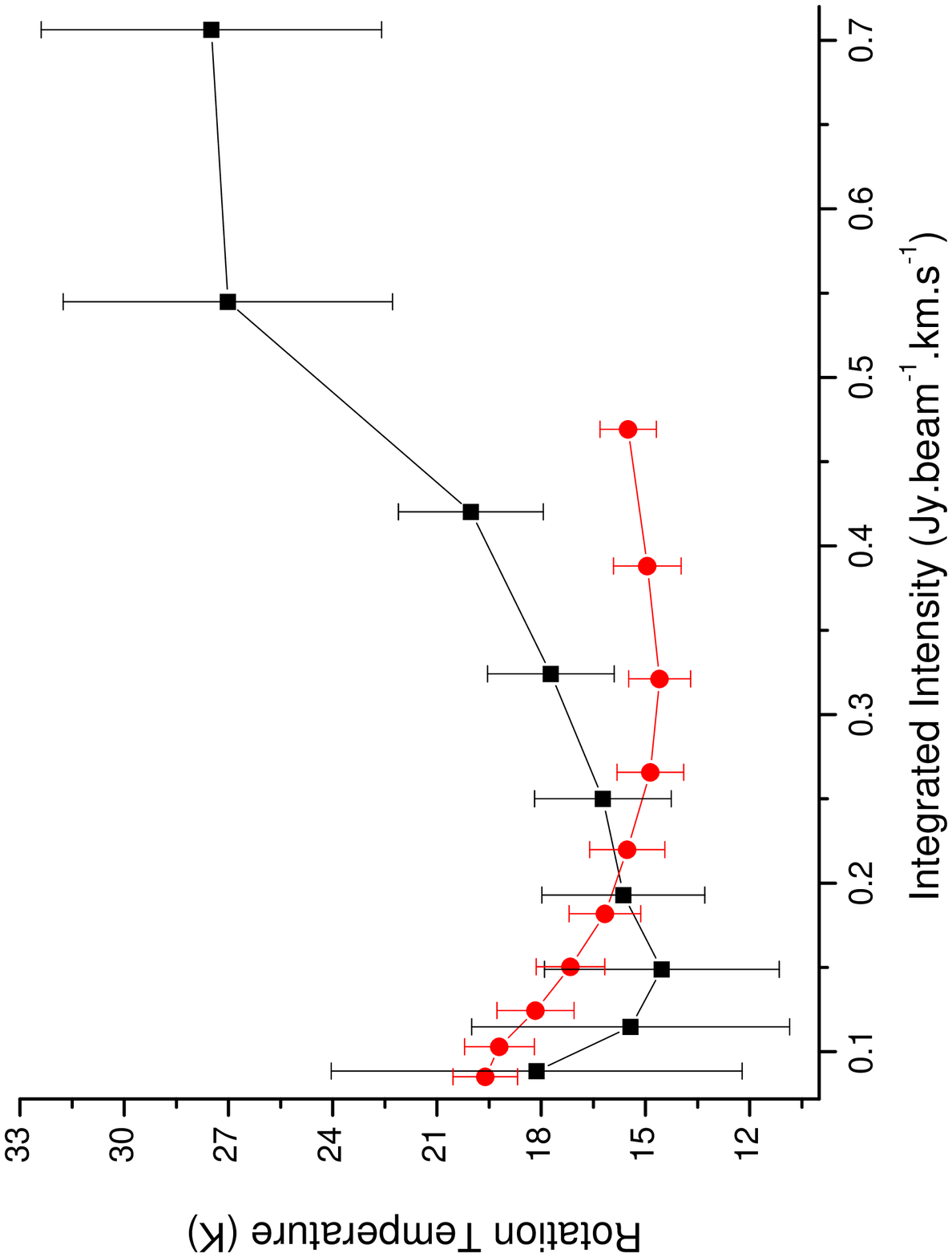}

%\special{psfile=f3.ps angle=-90 hscale=37 vscale=35 hoffset=220
%voffset=-60}

\caption{Relation between rotation temperature and \nh3 (1,1)
integrated intensity in the Northern Region
(black) and the Southern Region (red) of G28.34+0.06. The error bars 
present the $1 \sigma$ standard deviation.}

\end{figure}

\vskip 3in

\begin{figure}[h]
%\begin{center}

\includegraphics{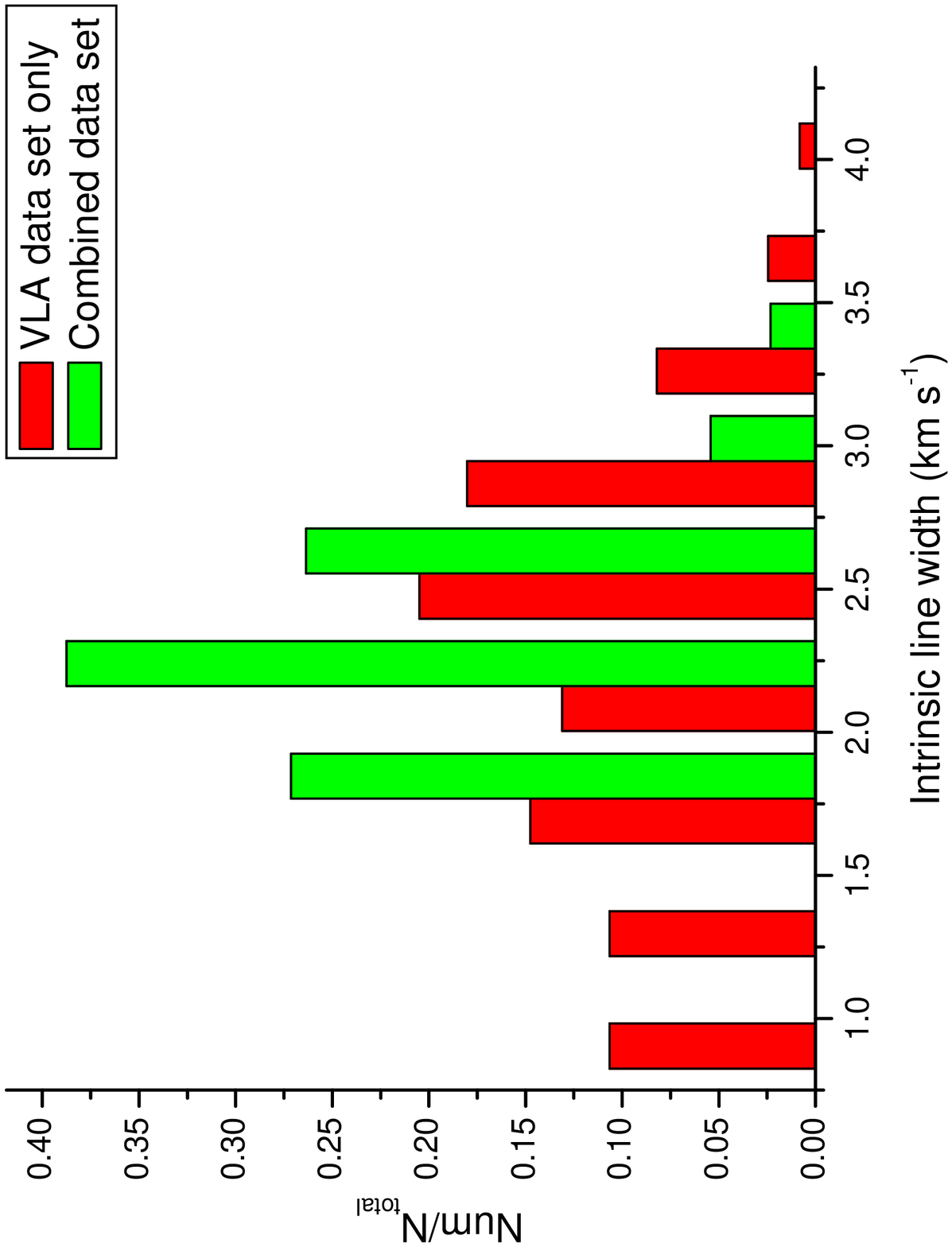}

\includegraphics{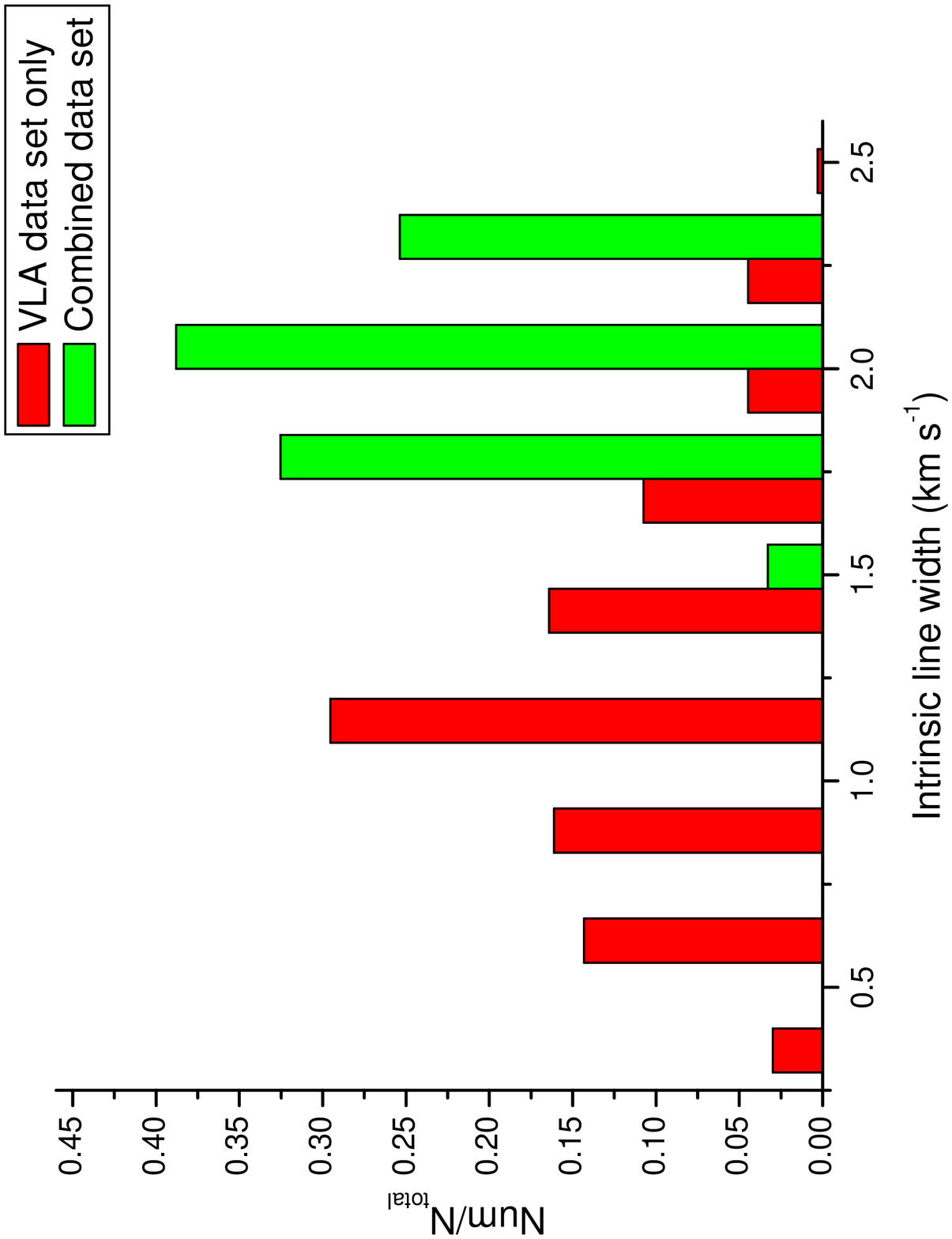}

\caption{ \nh3 (1,1) line width distribution in different velocity
bins toward G28.34+0.06. Red columns are the number count from the
VLA data set. Green columns are the number count from the combined
data set (VLA+100m). \textbf{Left:} Line width distribution for
clump P2 over an area of $10'' \times 10''$. \textbf{Right:} Line
width distribution for clump P1 over an area of $15'' \times
5''$. The \nh3 line widths, measured in FWHM, are corrected for optical
depth and broadening due to hyperfine structures.}

\end{figure}

\newpage


\begin{thebibliography}{}

%\bibitem[]{} Bally, John; Stark, Antony A.; Wilson, Robert W.; Langer, William
%D. 1987, ApJ, 312, 45

\bibitem[]{} Beuther, H.; Sridharan, T. K.,; Saito, M. 2005, ApJ, 634, L185

\bibitem[]{} Carey, S. J., Clark, F. O., Egan, M. P., et al. 1998, ApJ, 508,
721

\bibitem[]{} Carey, S. J., Feldman, P. A., Redman, R. O., et al. 2000, ApJ,
543, L157

\bibitem[]{} Castets, A., Langer, W. D., 1995, A\&A, 294, 835

\bibitem[]{} Cesaroni, R., Wilson, T. L. 1994, A\&A, 281, 209

\bibitem[]{} Di Francesco, J, Andre, P, Myers, P. C., 2004, ApJ, 617, 425

\bibitem[]{} Egan,M. P., Shipman, R. F., Price, S. D., et al. A Population of
Cold Cores in the Galactic Plane. 1998, ApJ, 494, L199

\bibitem[]{} Estalella, R., Mauersberger, R., Torrelles, J. M., Anglada, G., Gomez, J. F., Lopez, R. \& Muders, D.  1993, ApJ, 419, 698


\bibitem[]{} Hennebelle, P., Perault, M., Teyssier, D., Ganesh, S, 2001, A\&A,
365, 598

\bibitem[]{} Hillenbrand, Lynne A., 1997, AJ, 113, 1733

\bibitem[]{} Johnstone, Doug; Bally, John. 1999, ApJ, 510, 49

\bibitem[]{} Johnstone, Doug; Fiege, Jason D.; Redman, R. O.; Feldman, P. A.;
Carey, Sean J., 2003, ApJ, 588, 37

\bibitem[]{} Klessen, Ralf S., 2004, Ap\&SS, 292, 215

\bibitem[]{} Lada, Charles J., Lada, Elizabeth A., 2003, ARA\&A, 41, 57

\bibitem[]{} Larson, R. B. 1985, MNRAS, 214, 379

\bibitem[]{} Lis, D. C., Serabyn, E., Keene, Jocelyn, Dowell, C. D., Benford, D. J., Phillips, T. G., Hunter, T. R., Wang,
N., 1998, ApJ, 509, 299

%\bibitem[]{} McKee, Christopher F., Zweibel, Ellen G. 1995, ApJ, 440, 686

%\bibitem[]{} Motte, F., Andre, P., Neri, R., 1998, A\&A, 336, 150

\bibitem[]{} MacLow, Mordecai-Mark, 2004, Ap\&SS, 289, 323

\bibitem[]{} Mouschovias, T. C. 1991, ApJ, 373, 169

\bibitem[]{} Myers, P. C., 1998, ApJ, 496, L109

\bibitem[]{} Pillai, T., Wyrowski, F., Carey, S.J., \& Menten, K.M., 2006a,
A\&A, 447, 929

\bibitem[]{} Pillai, T., Wyrowski, F., Menten, K.M., \& E.Krugel, 2006b,
A\&A, 450, 569

%Preibisch, Th.; Ossenkopf, V.; Yorke, H. W.; Henning, Th. 1993,
%A\&A, 279, 577.

\bibitem[]{} Rathborne, J. M., Jackson, J. M., Chambers, E. T., Simon, R.,
Shipman, R., Frieswijk, W., 2005, ApJ, 630, L181

\bibitem[]{} Rathborne, J. M., Jackson, J. M., Simon, R., 2006, ApJ, 641, 389

\bibitem[]{} Rathborne, J. M., Simon, R., Jackson, J. M., 2007, ApJ, 662, 1082

\bibitem[]{} Redman, R. O., Feldman, P. A., Wyrowski, F., Cote, S., Carey, S.
J., Egan, M. P., 2003, ApJ, 586, 1127

\bibitem[]{} Shu, Frank H., Adams, Fred C., Lizano, Susana, 1987, ARA\&A, 25,
23

\bibitem[]{} Simon, R., Jackson, J. M., Rathborne, J. M., Chambers, E.
T., 2006, ApJ, 639, 227

\bibitem[]{} Teyssier, D., Hennebelle, P., P¨¦rault, M., 2002, A\&A, 382, 624

\bibitem[]{} Sridharan, T. K., Beuther, H., Saito, M., Wyrowski, F., Schilke,
P., 2005, ApJ, 634, L57

\bibitem[]{} Vogel, S. N., Bieging, J. H., Plambeck, R. L., Welch, W. J., Wright, M. C.
H., 1984, ApJ, 283, 655

\bibitem[]{} Ward-Thompson, Derek. 2002, Sci.,295, 76.

%\bibitem[]{} Williams, J. P., De Geus, E. J., \& Blitz, L. 1994, ApJ, 428, 693

%\bibitem[]{} Williams, J. P., Blitz, L. \& Stark, A. A. 1995, ApJ, 451, 252

\bibitem[]{} Wang, Y., Zhang, Q.,Rathborne, J. M.; Jackson, J \& Wu, Y., 2006, ApJ, 651, L125


\end{thebibliography}
\end{document}